# AN APPLIED STUDY ON EDUCATIONAL USE OF FACEBOOK AS A WEB 2.0 TOOL: THE SAMPLE LESSON OF COMPUTER NETWORKS AND COMMUNICATION

Murat Kayri[1] and Özlem Çakır[2]

[1]Department of Computer and Instructional Technology, Yuzuncu Yil University, Van, Turkey
mkayri@yyu.edu.tr
[2] Department of Computer and Instructional Technology, Ankara University, Ankara, Turkey
ocakir@education.ankara.edu.tr

### ABSTRACT

*The main aim of the research was to examine educational use of Facebook. The Computer Networks and Communication lesson was taken as the sample and the attitudes of the students included in the study group towards Facebook were measured in a semi-experimental setup. The students on Facebook platform were examined for about three months and they continued their education interactively in that virtual environment. After the-three-month-education period, observations for the students were reported and the attitudes of the students towards Facebook were measured by three different measurement tools. As a result, the attitudes of the students towards educational use of Facebook and their views were heterogeneous. When the average values of the group were examined, it was reported that the attitudes towards educational use of Facebook was above a moderate level. Therefore, it might be suggested that social networks in virtual environments provide continuity in life long learning.*

## 1. INTRODUCTION

Means of communication change together with improving technology. Today, the number of social networks, where communication is established, is rapidly increasing. As a part of daily life, mostly teenagers and adults use social networks such as Facebook, Myspace, Youtube, Weblogs, Xanga, Friendster, Orkut, Bebo and Wiki to take advantage of opening their world to friends and introduce themselves to others. At the same time, such users share their photos and videos, become members of groups and they are also provided with many msn and e-mail possibilities.

Facebook was started by the students of Harvard University in 2004 for only the students of the university. Later, it spread to other universities and gradually became a public domain [1]. Facebook, whose users are rapidly increasing, is the second largest social network in the world and the largest one in Turkey [2]. In Turkey, it is known that Facebook is generally used for the following reasons: finding friends, supervision, video, picture, photo, music and idea sharing, games, organizations, political reasons, e-trade, sexual reasons, and denouncement [2]. It is also reported that such social networks, as a means of communication, could be used for education [3, 4].

Boyd (2003) defines social networks as software products developed to make mutual interaction between individuals and groups easier, provide various options for social feedback and support the establishment of social relationships [5]. The features of social networks are summarized as follows [6, 7, 8, 9, 10, 11, 12, 13, 14]:

- Most social networks provide users with information sharing facilities such as e-mail, chats, instant messages, videos, blogging, file sharing and photo sharing,
- Social networks have a database for users so that they can easily find friends, form groups and share things with those with similar interests,
- Social networks provide users with opportunities for establishing on-line profiles and setup their own social networks,
- Most social networks are free of charge,
- Most social networks are reviewed and recreated according to user feedback. Similarly, open source versions enable users to develop their own applications to be integrated with sites,
- Social networks enable users to reset their own access and privacy; therefore users decide what to share and to what extent they share,
- Social networks focus on individual based personal online groups rather than first-generation online groups based on content, subjects or personal interests,
- Social networks allow constantly accumulating data update,
- Social networks allow data analysis,
- Social networks create a cooperative environment,
- Social networks support active model participation role through social features and chat facilities,
- Social networks provide interaction,
- Social networks provide users with a critical thinking environment,
- Social networks support active learning,
- Social networks provide school-student interaction and student-student interaction,
- Social networks increase student satisfaction with lessons,
- Social networks improve student writing skills,
- Social networks support informal learning Social networks allow optional profiles open to everyone,
- Social networks allow contact lists,
- Social networks allow users to follow connections between contact lists and other users in the system.

It is obvious that social network applications are closely related to many pedagogical points of constructivist approach, because of the above mentioned features [11]. They also provide people with individualized, personal settings. It is thought that the number of studies on social network applications in education is low, and further studies and research on educational use of such tools are suggested, since the previous ones focused rather on identification, network structure, privacy and technology [4].

In a study by [15], a project by the students of Queensland University of Technology, Department of Business Administration was introduced. On account of the fact that the students constantly used Facebook, a group page to enable those students to assess experience and proceeding during the course was created. With the study, it was concluded that Facebook could be used as a supplemental tool in education as a result of the students' digital proceeding and participation by the students.

In a study, Genç (2010) presented reflections on educational use of the following applications introduced by Web 2.0 technology: Weblogs, podcast and video casts, wikis, social networks, bookmarks, tagging and photo sharing. The study included details obtained during three undergraduate and one post-graduate courses offered by Firat University, Department of Computer Education and Instructional Technologies for a sample application of educational use of Facebook, a popular social network of the modern world, but the final reports were not presented [16].

Educational use of Facebook was thought to be an eligible subject for the study because of the following reasons: the number of social network users in intense communication was high; every user knew the setting and Facebook offered e-mail, forums and chats as a learning management system did. On the other hand, it was critical to reveal the reasons for daily intended use of the applications and rapid spreading since it could define possible factors to influence educational use.

As a result, the study presented an assessment of educational use of Facebook. To this end, the following sub-dimensions were examined:

1. Use of Facebook according to socio-demographic features
2. Levels of Facebook acceptance
3. Intended use of Facebook levels of educational use of Facebook

# 2. METHOD

## 2.1. Research Model

The study was a descriptive research and a general survey model was used, as the aim of the study was to show the attitude levels of the university students included in the study group towards Facebook application, a Web 2.0 tool, according to various sub-dimensions. At the same time, the study was a correlational research, since correlations between variables were taken into account. Also, it constituted a semi-experimental design because the study deliberately gathered the students on Facebook during one term (3 months) and the researchers observed the students in the setting to measure their attitudes towards use of Facebook.

## 2.2. Study Group

The study group consisted of 3$^{rd}$ graders of Ankara University, Faculty of Educational Sciences, Department of Computer Education and Instructional Technologies (CEIT) and Yuzuncu Yil University, Faculty of Education, Department of Computer Education and Instructional Technologies (CEIT) in 2008-2009 academic year, from Turkey. 31 students from Ankara University (18 male + 13 female) and 27 students from Yuzuncu Yil University (22 male + 5 female) were included in the study group. Mean age of the students in the study group was 22, 65±1, 59. The attitudes of the group towards Facebook were measured after interactively conducting "Computer Networks and Communication" lesson on Facebook during one term. Theory lessons (two hours) were conducted as scheduled by instructors in classroom settings and practice lessons were conducted with active participation of the students in a shared forum (group) on Facebook, a Web 2.0 tool. Information sharing by the students on Facebook was not restricted to the practice hours; the lesson was conducted with all kinds of activity at other times, on a voluntary basis.

## 2.3. Data Gathering Tool

The students in the study group were given three scales and a personal information form. The aim of the study was to show the attitudes of the students towards educational use of Facebook, a Web 2.0 tool. Their attitudes towards Facebook were measured by i) Facebook Adoption Scale ii) Intended Use of Facebook Scale and iii) Educational Use of Facebook Scale. Moreover, demographic data of the students in the study group was gathered by Personal Information Form. In the form, there were variables such as gender, age, and personal computer (PC) ownership, frequency of Facebook use, the amount of time spent on Facebook and group memberships on Facebook. The students in the study group were described by the variables in Personal Information Form and at the same time, correlation between the variables and the scores obtained by the measurement tools was examined. Therefore, the effects of the variables in Personal Information Form on the attitudes towards Facebook were revealed.

*i) Facebook Adoption Scale (FAS):*

"FAS", developed by [17], consisted of five sub-factors: "benefit", "ease of use", "social effect", "facilitating factors" and "community identity" and total 21 items. Cronbach Alpha reliability coefficient was calculated as .90. Scale rating 1-10 was used for the scale questions and the answers ranged from 1="I totally disagree" to 10="I totally agree.

*ii) Intended Use of Facebook Scale (IUFS):*

The scale developed by [18] to measure the intended use of Facebook, taking functions of Facebook into account, consisted of three sub-factors: "social relationships", "use for studies" and "daily use". The scale questions were five-Likert type and the answers ranged from 1="Never" to 5="Always. Cronbach Alpha reliability coefficient of the scale was 0.79.

*iii) Educational Use of Facebook Scale (EUFS):*

The scale developed by [18] consisted of three factors: "Communication", "Collaboration" and "Resource and Material Sharing". Likert type scale rating 1-10 was used for the scale questions and the answers ranged from 1="I totally disagree" to 10="I totally agree. The overall scale consisted of 11 items. Cronbach Alpha reliability coefficient of the scale was found as 0.938.

In the study, reliability analysis of the measurement tools was reexamined and Cronbach Alpha reliability coefficient of "Facebook Adoption Scale" was found as 0,889. It was found as 0,898 for "Intended Use of Facebook Scale" and finally as 0,937 for "Educational Use of Facebook Scale". The values of reliability coefficients indicated that the items of the measurement tool measured the attitudes towards Facebook with the minimum total error.

### 2.4. Procedure

Some assumptions were tested to determine statistical methods to be applied to total scores by the measurement tools given to the group. Normality of the obtained scores was tested by Kolmogorov-Smirnov Test. As a result of analysis, it was seen that total scale scores met the assumption of normality ($p>0.05$). Furthermore, total scores by the measurement tool met the assumption of homogeneity by Levene's statistics ($p>0.05$). Parametric methods were used for both descriptive statistics and hypothesis tests as total scores by the measurement tools were interval, and the assumptions of normality and homogeneity were met by the scales. However, non-parametric Spearman rho method was sometimes applied for correlation analysis in the study, because total scores and the correlated variables were occasionally discontinuous. Pearson Correlation Analysis, a parametric method, was used when total scores by the measurement tools and the other correlated variable were scale–continuous.

For all the statistical procedures, SPSS 16.0 was used.

## 3. FINDINGS

In this section, first of all, the variables in Personal Information Form are introduced. Table 1 shows "PC ownership" of the students included in the study group.

Table 1. Descriptive statistics of students' computer ownership variable

| Have you a computer? | Frequency | % |
|---|---|---|
| Yes | 55 | 94,8 |
| No | 3 | 5,2 |
| **Total** | 58 | 100 |

As it is clear from Table 1, most of the students included in the study group (94,8) had their own personal computers. The percentage was normal for the students from Department of Computer Education and Instructional Technologies and the fact that only three students (5,2%) did not have personal computers was caused by economic disadvantages.

The students were asked how long they had used the Internet. Table 2 presents feedback from Personal Information Form.

Table 2. Descriptive findings about the variable of "How many years have you used the Internet?"

| | N | Min | Max | Mean | Std.Deviation |
|---|---|---|---|---|---|
| **Years of internet use** | 58 | 3 | 12 | 6,67 | 2,40 |

According to Table 2, some of the students included in the study group reported they had used the Internet for at least three years and some for 12 years. The values was 6.67 ± 2,40, when the group mean was taken into consideration. They were asked about frequency of Facebook use in accordance with the purpose of the study, followed by data collection on Internet use period. In Table 3, the obtained findings are presented.

Table 3. Frequency of Facebook use

| Using Frequency | Frequency | % |
|---|---|---|
| Every day | 18 | 31,0 |
| A number of times in a week | 32 | 55,2 |
| A number of times in a week | 3 | 5,2 |
| A number of times in a year | 1 | 1,7 |
| Never use | 4 | 6,9 |
| **Total** | **58** | **100,0** |

31% of the students included in the study group used Facebook Web 2.0 every day (for any intended use) and most of them (55,2%) went online a few times a week. 6,9% of the students never used Facebook. The remaining students (5,2%+1,7%) reported they rarely used Facebook. The students included in the study group were then asked about the time spent on Facebook. In Table 4, the obtained findings are presented.

Table 4. Descriptive statistics of using Facebook time when surfed in Facebook

| Using time | Frequency | % |
|---|---|---|
| Less than 15 minutes | 13 | 22,4 |
| Half an hour (nearly) | 17 | 29,3 |
| Nearly 1 hour | 18 | 31,0 |
| Between 1-3 hours | 7 | 12,1 |
| More than 3 hours | 3 | 5,2 |
| **Total** | **58** | **100,0** |

As it is clear from Table 4, most of the students included in the study group were in the online range of 15 minutes and one hour. 12,1% of them actively used Facebook for 1 to 3 hours, and only a few spent more than 3 hours on Facebook.

It was reported that 96,6% of the students included in the study group were members of Facebook social groups. Levels of the students' attitudes will be presented following the introduction of the above mentioned variables.

In the study, the students were given three scales of Facebook and "Computer Networks and Communication" practice hours were conducted on Facebook during one term (three months). Then, the attitudes of the students towards Facebook acceptance, intended use and educational use were measured. The main aim of the study was to show the attitudes of the students towards educational use of Facebook. However, the attitudes towards Facebook acceptance and intended use as supporting variables were examined.

Table 5. Descriptive statistics related to measurement tools

| Measurement Tools | N | Min. | Max. | Mean | Std. Deviation |
|---|---|---|---|---|---|
| FAS | 58 | 73,00 | 214,00 | 142,79 | 32,04 |
| IUFS | 57 | 14,00 | 60,00 | 35,17 | 10,38 |
| EUFS | 57 | 22,00 | 106,00 | 73,19 | 19,86 |

When the statistics of the score variable from Facebook Acceptance Scale (FAS) were examined, it was seen that the lowest scale score was 73 and the highest was 214. The mean was 142,79 ± 32,04. As it is known, FAS consisted of 22 items and the highest scale score was 220. The highest scale score showed the maximum level of attitude. As the scale score increased, the acceptance level increased, as well. When the mean value in Table 5 (142,79) was examined, it was clear that the students included in the study group did not have high attitudes, but their attitudes were above average. When standard deviation value of FAS was examined, it was observed that the acceptance level was heterogeneous; and some of the students included in the study group had moderate attitudes and some had a tendency to have acceptance levels above average. "Benefit", "Ease of Use", "Social Effect", "Facilitating Factors" and "Community Identity" sub-dimensions of FAS were not separately examined. The acceptance levels in a general sense were shown under a single dimension. In short, we must not forget the fact that acceptance of Facebook, with an increasing number of users, has changed in a short term and thus it is doubtful that the obtained findings will not apply for a long period.

Intended Use of Facebook Scale (IUFS), taken as a supporting variable in the study, consisted of three sub-dimensions: "Social Relationships", "Use for studies" and "Daily Use". A high score from the measurement tool showed that Facebook was preferred for social relationships, academic studies and daily use. The highest score form the scale which consisted of 12 items was 60. It was determined that the lowest scale score of the students included in the study group was 14 and the highest was 60. The mean of the study group was found as 35,17 and standard deviation as 10,38. When the group mean was examined, it was remarkable that the levels of the students' attitudes in IUFS were parallel to those in FAS. IUFS was considered as slightly above average with a mean of 35,17. The standard deviation value of the group indicated a heterogeneous structure. It was concluded that some of the students thought Facebook could be preferred for social relationships, studies and daily used and vice versa. In the light of a normal distribution of the measurement tools (according to Kolmogorov-Smirnov Test), it was presumed that the number of the students with negative attitudes was similar to the number of the students with positive attitudes. In this context, it was concluded that the lesson conducted on Facebook during one term did not elevate the acceptance levels and the levels of internalized intended use.

Educational Use of Facebook Scale (EUFS), the main variable of the study, was examined. The scale consisted of three sub-dimensions and total 11 items: "Communication (six items)", "Collaboration (three items)" and "Resource and Material Sharing (two items)". When descriptive statistics of the study group were examined, the lowest score of the group was 22 and the highest 106. Since 1-10 scale rating was used for the scale, the minimum score was 11 and the maximum 110. As it is clear from Table 5, the mean EUFS of the study group was 73,19 ± 19,86. The students reported that Facebook could beneficially be used in education, with attitudes above average. As it was observed in the other measurement tools, EUFS had a heterogeneous structure caused by the standard deviation value of the attitude level. Since EUFS was the main component of the study, the sub-dimensions of the scale were separately examined. Table 6 presents the descriptive findings about EUFS sub-dimensions.

Table 6. Descriptive statistics related to sub-dimensions of EUFS

| **Factors** | **N** | **Min** | **Max** | **Mean** | **Std.Deviation** |
|---|---|---|---|---|---|
| Communication | 57 | 11 | 60 | 40,29 | 12,16 |
| Collaboration | 57 | 6 | 30 | 21,01 | 5,68 |
| Resource and material sharing | 57 | 4 | 20 | 11,87 | 4,10 |

There were six items of the factors of "Collaboration" of EUFS and the factors scores ranged from 6 to 60. As it is clear form the values of "Communication" sub-dimension presented in Table 6, the lowest observed group score was 11 and the highest was 60. When the mean value was examined, the contribution of EUFS to "Communication" was considered above average. The same applied to "Collaboration", represented by three items and "Resource and Material Sharing", consisting of two items.

Thus, the attitudes of the students included in the study group were average and above average in the three measurement tools. The study attempted to examine whether there was a correlation between Facebook acceptance, intended use and educational use levels and the variables (gender, age, how many years have they used the Internet? and etc.) in Personal Information Form. The analysis findings are summarized in Table 7.

Table 7. The correlation between students' attitude level and students' demographic variables

| Measurement tool/Variables | Correlation Method | R | P |
|---|---|---|---|
| FAS – Gender | Spearman rho | -0,07 | 0,960 |
| FAS – Facebook Using Frequency | Spearman rho | -0,216 | 0,103 |
| FAS – Years of Internet Using | Pearson | 0,089 | 0,505 |
| **FAS – Surfed time in Facebook** | **Spearman rho** | **0,304*** | **0,021** |
| | | | |
| IUFS – Gender | Spearman rho | -0,037 | 0,786 |
| IUFS – Facebook Using Frequency | **Spearman rho** | **-0,288*** | **0,030** |
| IUFS – Years of Internet Using | Pearson | 0,005 | 0,971 |
| IUFS **– Surfed time in Facebook** | **Spearman rho** | **0,313*** | **0,018** |
| | | | |
| EUFS – Gender | Spearman rho | | |
| EUFS – Facebook Using Frequency | Spearman rho | -0,069 | 0,609 |
| EUFS – Years of Internet Using | Pearson | 0,072 | 0,592 |
| EUFS **– Surfed time in Facebook** | **Spearman rho** | **0,373**** | **0,004** |

*: 0,05 significance level **: 0,01 significance level

As it is clear from Table 7, there was a significant correlation between Facebook adoption and the amount of time spent on Facebook (Spearman rho = 0,304; p<0.05). It was shown that Facebook adoption was not correlated with the other variables in Personal Information Form of FAS (p>0.05).

It was seen that there was a significant correlation between the IUFS attitude levels of the students included in the study group and "Frequency of Facebook Use" (Spearman Rho = -0,288; p<0,05). As frequency of Facebook use increased, the IUFS attitudes increased, as well. Here, the fact that coefficient r was minus was caused by the frequency of Facebook use variable was given as follows: 1- Every day, 2- A few times a week, 3- A few times a month and so on. In other words, as frequency categorically decreased, the attitude levels increased. In Table 7, the significant correlation between IUFS and the amount of time spent on Facebook is presented (Spearman rho =0,303; p<0.05). The other variables did not significantly correlate with IUFS (p>0,05).

When EUFS was examined, it was seen that there was a significant correlation between the attitude levels and the amount of time spent on Facebook (Spearman rho = 0,373; p<0,01). EUFS did not significantly correlate with the other variables.

## 3. CONCLUSIONS

In Turkey, Facebook, with an increasing number of users, is frequently preferred by students. In addition to the reports by [2], it might be suggested that Facebook can be used in education. It is natural that those who frequently use Facebook could easily use it in education, as well.

The fact that learning management systems like Facebook present e-mail, forums and chat facilities together with personal profiles enables instructors to use the systems without introducing an additional teaching management system. Also, according to changing network management system, Facebook must be improved. The emergence of next generation networks & services has ushered in a new era of technological advancement. At this time, the focus is to have some technology-independent, network-agnostic and completely autonomic management framework for networks and its related services [22].

In traditional classrooms, teacher talking time might be longer than student talking time [19]. This case is not experienced on Facebook. On the contrary, it is student centered. As lesson materials are saved on Facebook lesson page, students can access to information and answers to the previous questions by classmates do not prevent them from re-asking questions. In this respect, Facebook brings many educational advantages.

The study examined educational perceptions of university students about Facebook, a Web 2.0 tool. The attitudes of the students were measured by three different measurement tools within three months. The attitudes of the students included in the study group were strikingly heterogeneous. In correlational examination, it was seen that those who spent much time on Facebook perceived Facebook as an educational tool. In other words, those who had previously considered Facebook as a social setting had positive attitudes towards educational use of Facebook. It was shown that most of the students actively participated in virtual environment during the study, unlike the traditional method. It was observed that learning was shaped by the students, as constructivist approach suggested, and even lesson materials were developed by the students. In this respect, Facebook might be suggested as an effective learning environment.

In a study on educational use of Facebook [20], cognitive, affective and psychomotor abilities of students on Facebook were examined. It was discussed that when instructors who wanted to use Facebook in education were aware of the advantageous and disadvantageous aspects of technology-human interaction, Facebook provided rich lesson content and effective learning in a virtual environment was possible because of advantages like interactive communication. The same study reported that Facebook under no supervision might entail disadvantages. In the present study, it was recorded that the students were under control and became more participant when the instructors acted as moderators on Facebook. In this sense, Facebook management must be under the supervision of instructors.

In another study [21], student proceedings on Facebook were examined. The study found that when Facebook was preferred for educational purposes, proceedings in that virtual environment varied according to gender, ethnicity and educational background of parents. In the present study, it was concluded that gender was not influential on the attitudes towards Facebook, since the attitudes of the male and the female students included in the study group were similar.

Consequently, in case of considering some pedagogical issues, Facebook Web 2.0 tool can be used for educational purpose. Facebook media not only makes lesson enjoyable but also provides lots of electronic material. Building social network with Facebook provides collaboration in group.

**Dr. Murat Kayri**

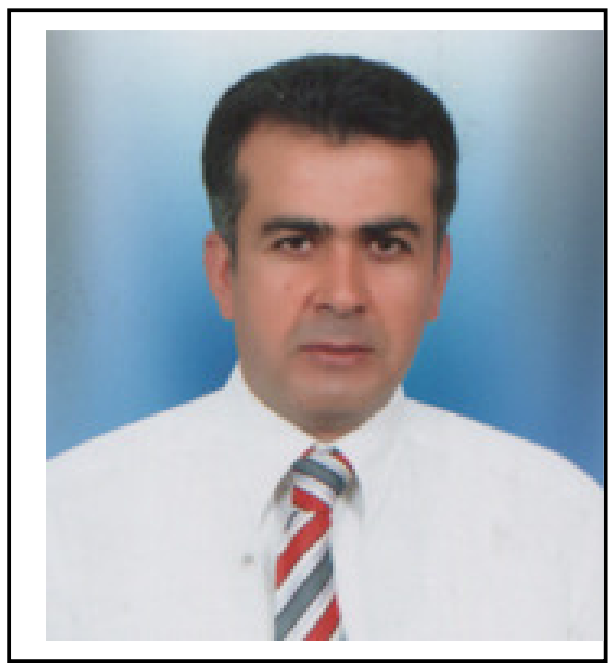

Dr. Kayri is an assistant professor in Computer Science and Instructional Technology Department in Yuzuncu Yil University. Dr. Kayri interests in neural network, statistical modelling, and networking. He has lots of articles on statistical and artificial neural network.

**Dr. Özlem Çakır**

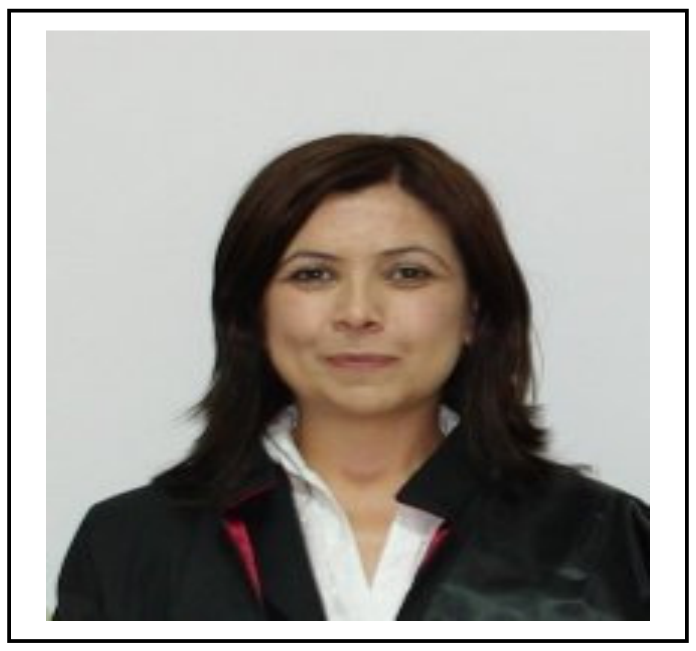

Dr. Çakır is a lecturer in Ankara University, in Department of Computer Science and Instructional Technology. She studies on social network, Web 2.0 tools and computer assisted learning. She also studies on internet dependency.